\documentstyle[aps,manuscript,epsf]{revtex}

\newcommand{\be}{\begin{equation}}
\newcommand{\ee}{\end{equation}}
\newcommand{\r}[1]{(\ref{#1})}

\def\fun#1#2{\lower3.6pt\vbox{\baselineskip0pt\lineskip.9pt
\ialign{$\mathsurround=0pt#1\hfil##\hfil$\crcr#2\crcr\sim\crcr}}}

\begin{document}

\def\baselinestretch{1.5}
\normalsize


\title{On the interactions of Skyrmions with domain walls }
\author
{A.E. Kudryavtsev \footnote{On leave from ITEP, Moscow, Russia and also from
IKP, Jelich, Germany. 
e-mail:kudryavt@heron.itep.ru},
B.M.A.G. Piette \footnote {e-mail:B.M.A.G.Piette@uk.ac.durham}
and
W.J. Zakrzewski \footnote {e-mail:
W.J.Zakrzewski@uk.ac.durham}}
\address{Department of Mathematical Sciences, University of Durham, Durham, 
~Durham DH1 3LE,GB}

\maketitle

\begin{abstract}

We study classical solutions of a particular version 
of the modified Skyrme model in (3+1) dimensions. The model 
possesses Skyrmion solutions as well as stable domain walls that connect 
different vacua of the theory. We show that there is an attractive 
interaction between Skyrmions and domain walls.
Thus Skyrmions can be captured by the domain walls. 
We show also that, when the mass term is of a special type, 
the model possesses bound states of Skyrmions and of the domain wall. 
They look like deformed 2-dimensional Skyrmions captured by the 
wall. The field configurations of these solutions can interpreted as having
come from the evolution of the  3-dimensional Skyrmions 
captured  by the domain wall. For more conventional choices of the mass term 
of the model in the model the attraction between the Skyrmions and the wall
leads to the capture of the Skyrmions which are then turned into
topological waves which spread out on the wall.
 We have observed, numerically, such captures and the emission of the waves.
  We speculate that this observation  may be useful in the explanation 
of the problem of baryogenesis and  baryon-antibaryon  
asymmetry of  the Universe.


\end{abstract}

\def\baselinestretch{1.5}
\normalsize
\vspace{2mm}
\vspace{2mm}

\section{Introduction}

The Skyrme model in (3+1) dimensions, originally introduced by Skyrme
\cite{1}, has attracted a lot of attention as a reasonable good effective 
theory for
 describing baryons \cite{2}. Baryons in this  theory are
described by  topologically nontrivial classical configurations (Skyrmions). 
Recently there has been some discussion of the problem 
of the baryon-antibaryon asymmetry of the Universe. 
This problem may be related to the 
problem of interaction between Skyrmions (or baryons) and
 domain walls during the early stages 
of the Universe.

The interaction of particles with domain walls has become a
 subject of interest following a seminal paper by Voloshin \cite{3}.
 A detailed analysis of the  scattering 
of Abelian gauge particles on domain walls may be 
found in a paper by Farrar et al 
\cite{4}. The problem of  the chiral fermion determinant in the presence of a 
domain 
wall was discussed in \cite {5}. It is also worth mentioning that the 
expanding bubbles 
of a new phase during the electroweak phase transition may be an additional 
source for 
the CP - violationg effects of the electroweak baryogenesis, see, {\it e.g.}
\cite {6,7}.

Recent interest in the problem of the interaction of particles with
 domain walls was generated
by a paper of Dvali et. al \cite{8}. In this paper the authors consider the 
interaction of monopoles with domain walls. They suggest
 that this interaction may help to solve the problem of low density of
 monopoles in the Universe. They conjecture
that the domain walls would sweep up all the monopoles
which would then defuse along the domain walls. 

Very recently this 
conjectured interaction of monopoles with domain walls was  
supported by numerical stimulations of monopole-domain wall collision processes
as reported in \cite{9}.

The fact that extended objects, such as Skyrmions, interact nontrivially with 
domain walls had already been discussed in our previous paper \cite{10}. 
In that paper we studied the 
interaction of Skyrmions with domain walls in a two-dimensional baby-Skyrmion 
model \cite{11}. In \cite{10} we showed that domain walls can attract and 
absorb Skyrmions. 
After its collision with a domain wall the Skyrmion splits into two parts
 which propagate (with the speed of light) along 
the domain wall. Each part of the former Skyrmion looks like a 
wave packet which carries one half of the topological charge of 
the Skyrmion. That is  why the 
compound objects like Skyrmions and  monopoles 
can interact nontrivially with domain walls. 
This interaction requires further careful study.

In this paper we concentrate our attention on the problem of the interaction 
of Skyrmions with domain walls in a realistic 3-dimensional case. 
To do this we consider the conventional  Skyrme model \cite{1,2} but with a 
slightly modified expression for the mass term. Our modification is motivated 
by the existence of solutions in the form of domain walls. In fact, first we 
introduce the simplest modification which is then further altered by the 
introduction of an additional parameter.
The model is introduced and discussed in detail in the next 
section. For our choice of the mass term it possesses solutions
which correspond to both Skyrmions and domain walls. Next we study the 
problem of their interaction. We demonstrate that there exists a solution, 
which may be considered as a bound state of a Skyrmion and a domain wall.
We also show that the Skyrmions and the domain walls attract each other. 
We have performed a series of numerical simulations studying this interaction. 
In our simulations we have always found the attraction although
the final stage of the evolution process({\it ie} whether the topological 
waves on the wall spread out or oscillate)  depends on the 
details of the modified mass term of the model. In each case the
Skyrmions are captured, resulting in the emission of topological
waves. The waves spread out on the wall in the model with
the simplest version of the modified mass term, or oscillate when the mass
term is further modified to prevent this spreading.

\section{The Model}

To study the interaction between Skyrmions and domain walls in (3+1) 
dimensions we consider a model described by the Lagrangian density
\be
L=L_{kin}+L_{Sk}+ L^{mod}_{mass},
\label{EqLag}
\ee
where $L_{kin}$ and $L_{Sk}$ are the standard kinetic and Skyrme terms
 for the effective 
chiral field theory \cite{1,2} and so are given by:
\be
L_{kin}=\frac{F_{\pi}^2}{16} Tr ( U U^+),
\ee
\be
L_{Sk}= \frac{1}{32e^2} Tr [\partial_{\mu}UU^+,\partial_{\nu}U U^+]^2.
\ee
Here $U$ is an $SU(2) ~~ 2\times 2$  matrix, $U=\Phi_0+i \vec \tau  \vec \Phi$
                 and 
$\Phi_0$ and ${\vec \Phi}$
 are real scalar fields, which satisfy the constraint

\be
\ \Phi_0^2+ {\vec \Phi}^2=1.
\label{EqConstraint}
\ee

The third term,  the mass term, violates chiral symmetry and is very 
nonunique, as it is determined only in the limit of small pion fields. In our 
case we use the following form of the mass term:
\be
L_{mass}^{mod} = \frac{1}{64} m_{\pi}^2 F^2_{\pi} Tr [(U+U^+-2)(U+U^++2)].
\label{EqModMassTerm}
\ee
Our choice \r{EqModMassTerm} of the
 mass term gives us two different absolute minima for the energy 
of the field configuration at constant $U=\pm1$ (vacua states of the theory). 
Note that the more usual mass term leads to the existence of 
only one vacuum state of the theory 
(at $U=1$):
\be
\ L_{mass}= \frac{1}{8} m_{\pi}^2 F^2_{\pi} Tr ( U +U^+ -2).
\label{EqMassTerm}
\ee
Note also that when the field configuration differs slightly from 
the vacuum $U=1$, 
i.e. $|U-1|\ll1$, the two mass
terms \r{EqModMassTerm} and \r{EqMassTerm} are very similar.

Both the modified \r{EqModMassTerm} and the standard \r{EqMassTerm} versions of the  Lagrangian  
possess solutions of the Skyrmion type.
However, the  modified version 
of the model possesses also solutions in the form of domain walls. The 
explicit expression for this solution can be easily
obtained after a further 
change of variables from the $({\vec\Phi} ,\Phi_0)$ field variables. Namely,
 we replace the $({\vec\Phi}, \Phi_0)$ by
$(f,g,h)$ defined by
\be
\ ({\vec \Phi},\Phi_0)=(\sin f\sin g\sin h, \sin f \sin g \cos h, 
\sin f \cos g, \cos f),
\ee
where $ f=f(\vec r, t)$, $g=g(\vec r,t)$ and $h=h(\vec r,t),f\in[0,\pi],
g\in[0,\pi],
h\in[0,2\pi]$ . Note that the constraint \r{EqConstraint} 
is automatically satisfied. 
The Lagrangian \r{EqLag} , when written in terms of
$f,g$ and $h$ is given by:
\begin{eqnarray}
& {\cal L} = {F_{\pi}^2 \over 8}
\bigr\{ f_{\mu}^2 + \sin^2f \, g_{\mu}^2 + \sin^2 f\,\sin^2 g\,h_{\mu}^2
+ k\, \sin^2f\,[(f_{\mu} g_{\mu})^2 - f_{\mu}^2\, g_{\nu}^2 \nonumber\\
& +\sin^2g\,((f_{\mu} h_{\mu})^2 - f_{\mu}^2\, h_{\nu}^2 )
+ \sin^2f\, \sin^2g \,((h_{\mu} g_{\mu})^2 - g_{\mu}^2\, h_{\nu}^2)]
-2\,m_{\mu}^2\, V(f,g) \bigl\},
\end{eqnarray}
where $k = 2/F^2_{\pi}e^2$.

Consider now the case when fields $g$ and $h$ are constant.
 In  this case we get from \r{EqModMassTerm}
the following equation of motion for the field $f$:
\be
f_{\mu}\sp{\phantom{a}\mu}+\frac{m^2_{\pi}}{2}\sin2f=0.
\label{EqSinG}
\ee
Among solutions of this equation \r{EqSinG} there is the static soliton 
solution 
\be
f_{sol}(x) = 2\arctan \exp {(\pm m_{\pi}(x-x_0))}.
\ee

In the (3+1) dimensional space-time this solution looks like a domain wall 
which links the two vacuum states $U=\pm1$ of the theory.

In terms of the angular variable $f$ the Skyrmion solution becomes:
\be
U(\vec r)= \cos f_{Sk}(r) + i{\vec \tau}{\vec n}\sin f_{Sk}(r),
\ee
where $f_{Sk}$ is a Skyrmion 
profile function \cite{2}
satisfying the boundary conditions $f_{Sk}(0)=\pi$, $f_{Sk}(\infty)=0 $
and $\vec n=\vec r/r$. The Skyrmion solutions are characterized by an integer 
valued degree of  the mapping of $S^3$ (compactified $R^3$ physical space) 
into $S^3$ - isospace. The analytical formula for this degree of mapping 
({\it ie} the winding number), when written in terms of the
fields $f$, $g$ and $h$, is 

\be
Q=-6\int  \sin^2f\, \sin g\, \epsilon_{ijk}\, f_i\, g_j\, h_k dx^3,
\ee
where
we have used Latin indices to denote spatial variables.

\section{A topological soliton on the domain wall}

As it was shown in our previous paper \cite{10}, 
a (2+1) - dimensional Skyrme-like 
field theory possesses
solutions which correspond to topological waves. 
They behave like massless wave
packets which carry the topological charge. 
It would be interesting to see whether analogous 
solutions exist in a (3+1)-field theory. 

To answer this question let us seek solutions of the equation of motion 
for the Lagrangian \r{EqModMassTerm} in the form of a generalized
 cylindrically symmetric  hegehog ansatz:
\be
f=f(t,x,\rho),\quad g=g(t,x,\rho),\quad  h=n\varphi,
\label{Eqfgh}
\ee
where $n$ is integer and we have changed Euclidean coordinates $(x,y,z)$ to
 the cylindrical ones $(x,r,\varphi)$,
\be
(x,y,z) \to (x,\rho\cos\varphi,\rho\sin\varphi).
\ee
It is interesting to note that $h=n\varphi$ is compatible with the equations
of motion ({\it ie} the $h$ equation is automatically
satisfied). So we have to look at the remaining equations. In 
terms of functions 
$f$ and $g$ the Lagrangian becomes:
\begin{eqnarray}
&{\cal L} = {F_{\pi}^2 \over 8} \int
\bigr\{ f_t^2  -f_x^2 - f_\rho^2 + \sin^2f \, (g_t^2  -g_x^2 - g_\rho^2) 
- \sin^2 f\,\sin^2 g\,{n^2\over \rho^2}\nonumber\\
&+ k\, \sin^2f\,[(f_t\,g_x-f_x\,g_t)^2 +(f_t\,g_\rho-f_\rho\,g_t)^2 -
(f_x\,g_\rho-f_\rho\,g_x)^2 + 
 \sin^2g\,{n^2\over \rho^2} (f_t^2  -f_x^2 - f_\rho^2)\\
&+ \sin^2f\, \sin^2g \,{n^2\over \rho^2}\,(g_t^2  -g_x^2 - g_\rho^2 )]
-2\,m_{\mu}^2\, V(f,g) 
\bigl\} 2\pi\,\rho\,d\rho\,dx,
\label{EqLagfgh}
\end{eqnarray}

In \r{EqLagfgh} we have used an expression more general
than \r{EqModMassTerm} for the mass term. Our new mass term depends on 
both variables $f$ and $g$. The purpose of this generalization will become
obvious soon. 
The equations of motion for functions $f$ and $g$, which follow from the 
Lagrangian \r{EqLagfgh}, are:
\begin{eqnarray}
&f_{tt}-f_{xx}-f_{\rho\rho}-{f_\rho\over \rho} 
- \sin f\,\cos f\, (g_t^2 -g_x^2 - g_\rho^2  - \sin^2g\,{n^2\over \rho^2}) 
+  m_{\mu}^2\,{\partial V(f) \over \partial f}\nonumber\\
&+k\,\sin f\,\cos f\,\bigl[
 (f_t\,g_x-f_x\,g_t)^2 +(f_t\,g_\rho-f_\rho\,g_t)^2 -(f_x\,g_\rho-f_\rho\,g_x)^2
+ \sin^2g\,(f_t^2 -f_x^2 - f_\rho^2)\,{n^2\over \rho^2}\nonumber\\ 
&- 2\,\sin^2f\, \sin^2g \,(g_t^2 -g_x^2 - g_\rho^2)\, {n^2\over \rho^2}
\bigr]
+k\,\sin^2 f\,\bigl[
(f_{tt}(g_x^2+g_\rho^2) -(f_{\rho\rho}+{f_\rho\over \rho})(g_x^2-g_t^2)\nonumber\\ 
&- f_{xx}(g_\rho^2-g_t^2) -2 f_{xt} g_x g_t - 2 f_{\rho t} g_\rho g_t + 2 f_{x\rho} g_x g_\rho
+g_{t\rho} (f_t g_\rho + f_\rho g_t)  + g_{tx} (f_t g_x + f_x g_t)\nonumber\\
&-g_{x\rho}(g_x f_\rho + g_\rho f_x) -g_{tt}(f_\rho g_\rho + f_x g_x) 
- g_{xx} (f_t g_t - f_\rho g_\rho) - g_{\rho\rho} (f_t g_t - f_x g_x) 
+ {g_{\rho} \over \rho} (f_x g_x - f_t g_t)\nonumber
\\
&+ 2\,\sin g\, \cos g \,{n^2\over \rho^2} (f_t g_t -f_x g_x - f_\rho g_\rho)
+\sin^2g\,(f_{tt} -f_{xx} - f_{\rho\rho}+{f_\rho\over \rho})\,{n^2\over \rho^2}
\bigr] = 0.\nonumber
\end{eqnarray}

and

\begin{eqnarray}
&g_{tt}-g_{xx}-g_{\rho\rho}-{g_\rho\over \rho} 
+ 2\,\hbox{cot}f\,(f_t g_t -f_x g_x - f_\rho g_\rho) 
+ \sin g\,\cos g\,{n^2\over \rho^2}
+ {m_{\mu}^2\over \sin^2 f}\,{\partial V(f,g) \over \partial g}\nonumber\\
&+ k\, \bigl[ f_{t\rho} (g_t f_\rho + g_\rho f_t)
+ f_{tx} (g_t f_x + g_x f_t) 
-f_{x\rho}(f_x g_\rho + f_\rho g_x) - f_{tt} (g_\rho f_\rho + g_x f_x)\nonumber\\ 
&- f_{xx} (g_t f_t - g_\rho f_\rho) 
- f_{\rho\rho} (g_t f_t - g_x f_x) + {f_{\rho} \over \rho} (g_x f_x-g_t f_t)
-(g_{\rho\rho}+{g_\rho\over \rho})(f_x^2-f_t^2)
- g_{xx} (f_\rho^2-f_t^2) \nonumber\\
&+g_{tt}(f_x^2+f_\rho^2)- 2 g_{xt} f_x f_t - 2 g_{\rho t} f_\rho f_t + 
2 g_{x\rho} f_x f_\rho
+ \sin^2 f\,\sin^2 g\,(g_{tt}-g_{xx}-g_{\rho\rho}+{g_\rho\over \rho})\,{n^2\over \rho^2}\\
\label{Eqf}
&+ 4\,\sin f\,\cos f\,\sin^2g\,{n^2\over \rho^2}(f_t g_t -f_x g_x - f_\rho g_\rho)
+ \sin^2 f\,\sin g\, \cos g\, {n^2\over \rho^2} (g_t^2 -g_x^2 - g_\rho^2) \nonumber\\
&- \sin g\,\cos g\, {n^2\over \rho^2} (f_t^2 -f_x^2 - f_\rho^2)
\bigr] = 0\nonumber
\end{eqnarray}

It is worth emphasising here, that setting $h=n\varphi$ has lead to a 
serious simplification of the problem. In  fact, 
our choice of $h$ corresponds to using one integral of motion $n$, 
whose conservation reflects the internal symmetry of the solutions of
the theory \r{EqLag}/. Note that by introducing our hegehog ansatz
we have been able to reduce the problem to that of having to 
solve only two coupled equations for functions $f$ and $g$ in (2+1) dimensions 
instead of having to consider the  coupled partial 
differential equations for three functions $f$, $g$ and $h$ in (3+1) 
dimensions.

Note also that in our ansatz the topological charge becomes:
\be
Q= -12 \pi n \int \sin^2f \sin g (f_\rho g_x-f_xg_\rho)\,dx\, d\rho.
\ee

We note that this expression is nonzero only when $n\ne0$. Thus to check
whether the hegehog ansatz gives us a Skyrmion solution
we see that we have to find solutions
of equations \r{Eqf} with the  boundary conditions:
\be
f(-\infty,\rho)=0, f(+\infty,\rho)=\pi; g(x,0)=\pi, g(x,+\infty)=0.
\label{EqBC}
\ee
These boundary conditions guarantee a nontrivial topological charge of the 
system. Note that there are no nontrivial solutions of  
\r{Eqf} within the class of functions with separated 
variables, {\it i.e.} of the form 
\be
f=f(x), g=g(\rho).
\label{Eqfg}
\ee
 This can be easily verified by looking at the Lagrangian \r{EqLagfgh}, 
in which the term
$$
\sim {(f_xg_{\rho}-f_{\rho}g_x)^2}
$$
clearly indicates that it is not possible
to find any nontrivial solutions of the form \r{Eqfg}.

Let us look next for the stationary solutions of eqs. \r{Eqf}. Consider the 
expression for the mass term $V(f,g)$ of the form
\be
V(f,g)=\frac{1}{2} \sin^2f (1-\delta_{m}\cos^2g),
\ee
where $0<\delta<1$. Using this form of $V(f,g)$,
and performing a numerical simulation, we have managed to find a solution 
of equations \r{Eqf} with the boundary conditions \r{EqBC}. This solution looks like 
a  slightly deformed  2-dimensional   (in $\rho, \varphi$ variables) Skyrmion 
(or hegehog) located on the domain wall near the origin $x=0$. The picture of 
the fields $f$ and $g$ as well as the energy density for this solution is 
shown in Figure 1 using the $r,x$ coordinates. One sees clearly that it 
corresponds to a bound states of a Skyrmion and a domain wall. The 
topological charge of the configuration is 1.  

The solution is stable with respect to the 3-dimensional 
$ (\vec r\to \epsilon\vec r)$ scaling transformations  
 as well as with respect to 2-dimensional ones, 
$ (x,\rho, \varphi) \to (x, \lambda\rho, \varphi)$.
Notice that the stability condition with respect to these 2-dimensional 
transformations
is similar to the condition for Skyrmions in the 
 baby-Skyrmion model in (2+1) dimensions, where a mass term
is needed to guarantee their stability
(see, e.g. \cite{11} and references therein).

In the case when $\delta_m \to 0$ , {\it i.e.} for the case when 
\be
V(f,g)= \frac {1}{2} \sin^2f
\ee
any initially formed field configuration which carries
a topological charge located 
on the domain wall is unstable with
 respect to the scale transformations
 in the $(\rho, \varphi)$-plane. This is 
why any initially formed field 
configuration in the case $\delta_m=0$ looks like a 
wave packet, which spreads along the domain wall. This shows 
 that in the case of the
Lagrangian \r{EqLag} with the mass term given by \r{EqModMassTerm}
 nontrivial topological configurations on 
the domain wall manifest themselves as topological radial waves.  Thus 
the Skyrmion, located outside  the domain wall, when  captured, is  absorbed
by the domain wall. A necessary condition for this capture process is an
 attraction between the Skyrmion and the domain wall.

\section{Attraction between Skyrmions and domain walls}

In this section we show that Skyrmions can be absorbed by a domain wall
and form field configurations described in the previous section. 
To demonstrate this we have performed numerical simulations
of the time evolution of various field configurations
describing Skyrmions and domain walls. In our work we used the
 Lagrangian \r{EqLag} written in terms of  
three angular fields $f$, $g$  and $h$. We were forced to consider this, a 
more general configuration than \r{Eqfgh},  as for an isolated domain wall 
the field $h$ is constant but this is incompatible with the hegehog ansatz
which describes the Skyrmion. Thus, to study their
interaction we need to consider the most general fields.
We see that the problem of determining the interaction of 
Skyrmions and domain walls is 
more complicated than the problem of looking
for bound states of Skyrmions and domain walls. 

First we have performed the numerical evolution of a Skyrmion placed at a 
finite distance from the domain wall in the first modified model 
($\delta = 0$). We have done this for various relative orientations of the 
Skyrmion with respect to the domain wall. In eah case the evolution was 
similar for every
orientation that we tried: the Skyrmion was attracted by the wall to finally 
merge with it. The structure formed by this process was then dissipated 
by radiating concentric waves on the wall. Figure 2 shows the evolution of 
a cross section in the $x-y$ plane (the plane perpandicular to the wall) of 
the energy density for this process. One sees that, after beeing absorbed, the 
Skyrmion is transformed into a wave which happens to be topologically non 
trivial.

Next we have performed the same simulation when $\delta = 0.3$. Again the
Skyrmion was absorbed by the wall except that this time instead of being 
radiated along the wall, it formed a bound state with the wall.
In Figure 3 we show the evolution of a cross section in the $x-y$ plane 
of the energy density for this process. 
This configuration caries one unit of topological charge and is 
is very similar to the topological structure described in the previous section.

We then looked at the interaction between a 2-Skyrmion configuration and the 
wall using $\delta = 0.3$ again. As for the single Skyrmion, the two Skyrmions
where attracted by the wall and merged with it to form a static ring-like 
structure on the wall. In Figure 4, we show the cross section of the energy 
density in the  $x-y$ plane and in the $y-z$ plane (on the wall) after the 
structure had settled down.


\section {Summary and Final Comments}
We have established 
the existence of bound states of Skyrmions and domain walls.
Depending on the form of the  mass term in the theory this solution is either 
stable or unstable. In the case of $O(4)$-symmetry
broken down to $O(3)\times Z_2$ this solution is unstable and its evolution 
looks like an evolution 
of a wave packet on the domain wall which carries the topological charge.
We have also shown that 
Skyrmions can be absorbed by the domain wall 
and that their field configuration is transmuted into
topological waves on the wall.
Considering the topological charge as the baryonic charge, this picture 
corresponds to the absorption of baryons by the domain walls.
Note that, because of the  different 
vacua to the left and to the right of the domain wall,  Skyrmions and 
antiskyrmions (or baryons and antibaryons) behave differently. We end with the
speculation that these arguments, {\it ie} the separation of Skyrmions and 
antiskyrmions by the domain wall and the observation
of the absorbtion of Skyrmions by the walls may be 
relevant to the discussion of the observed baryon-antibaryon  asymmetry of the 
Universe.


\centerline{\large \bf Acknowledgments.}

\vspace{2mm}
  This work was partly supported by RFFI grants 96-15-96578, RFBR-98-02-17316
and by a Royal Society grant.

\begin{figure*}[tp]
\unitlength1cm\hfill\newline
\begin{picture}(8,8)
\epsfxsize=8cm
\epsffile{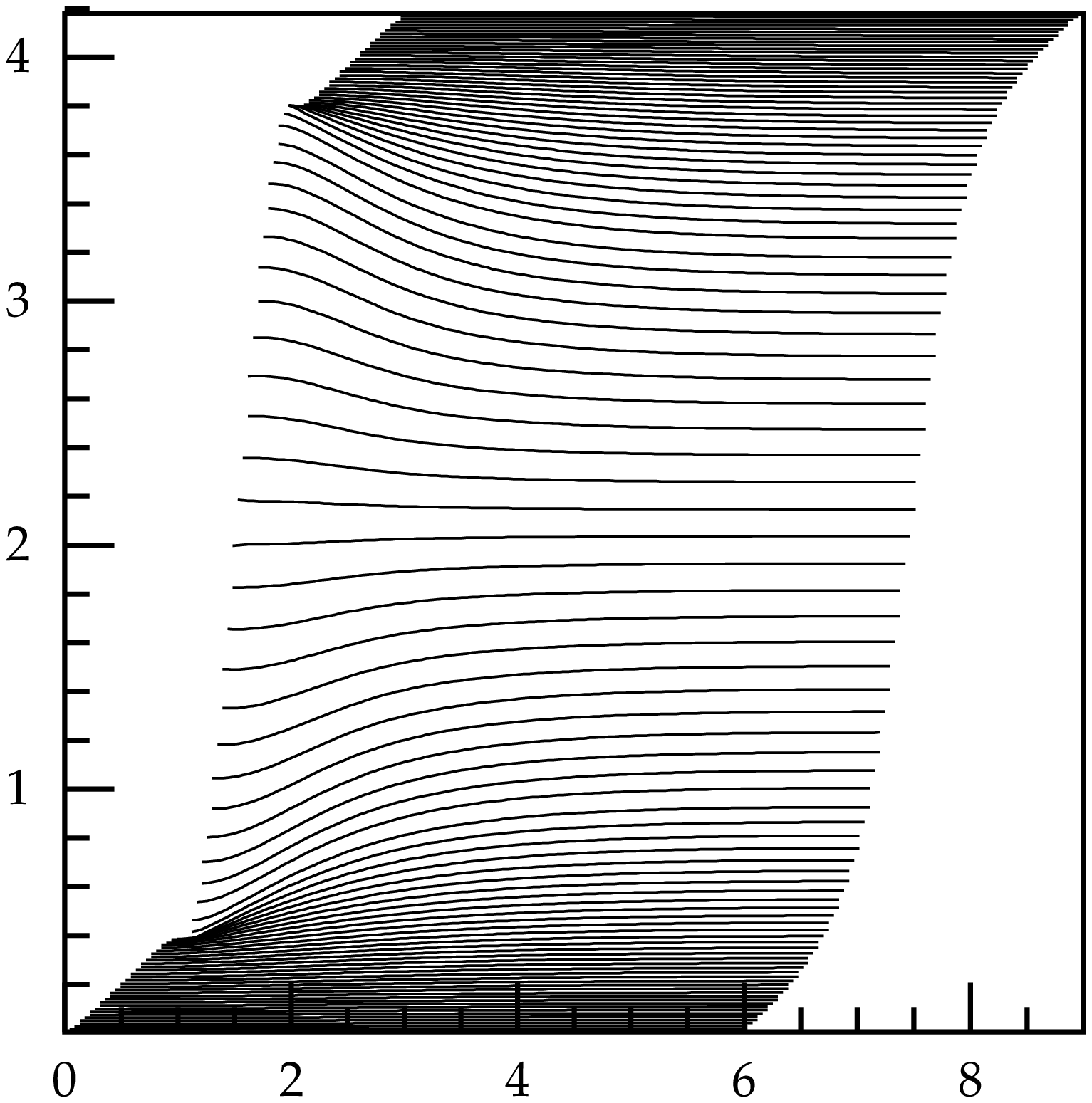}
\end{picture}
\begin{picture}(8,8)
\epsfxsize=8cm
\epsffile{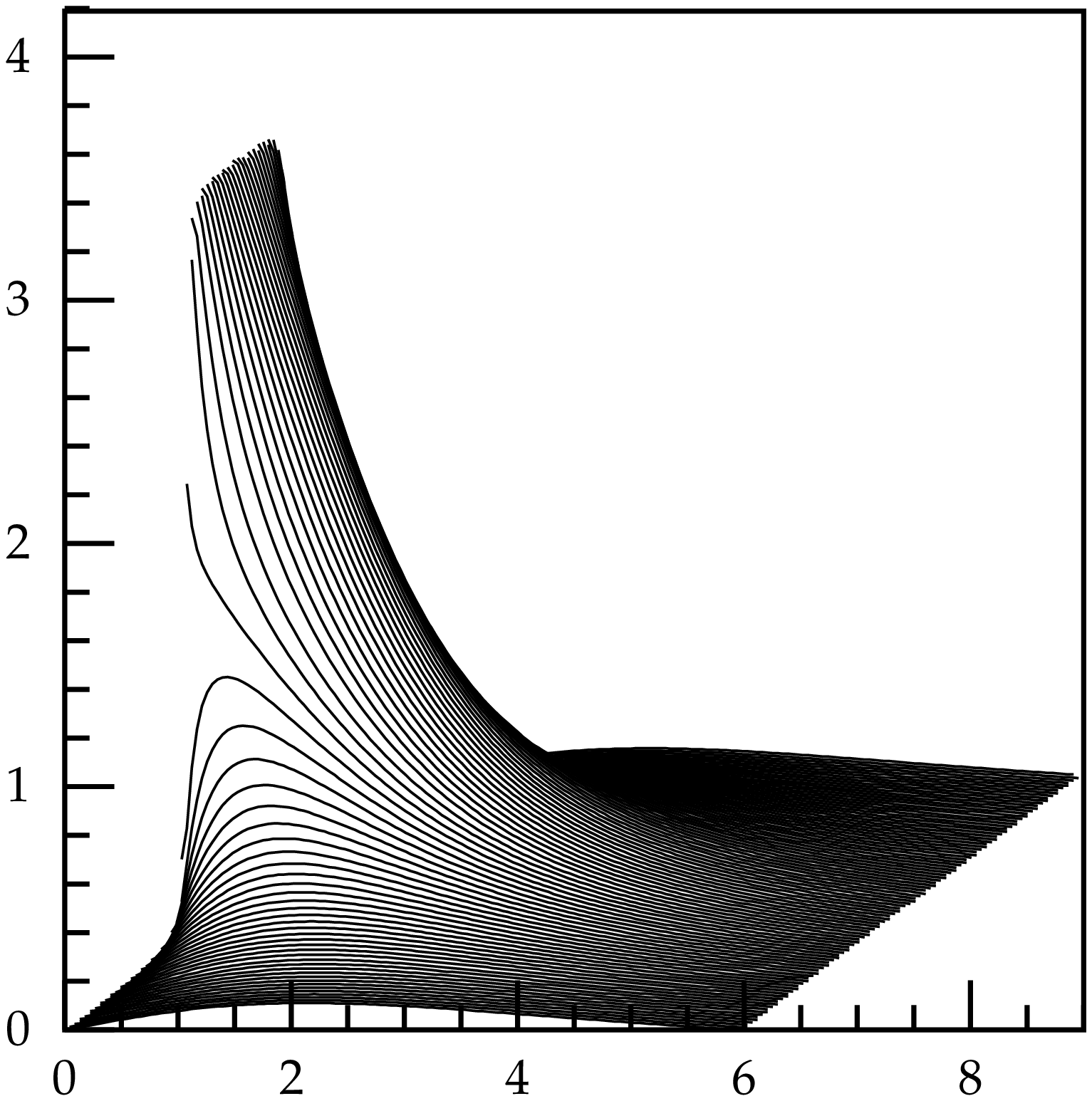}
\end{picture}
\begin{picture}(8,8)
\epsfxsize=8cm
\epsffile{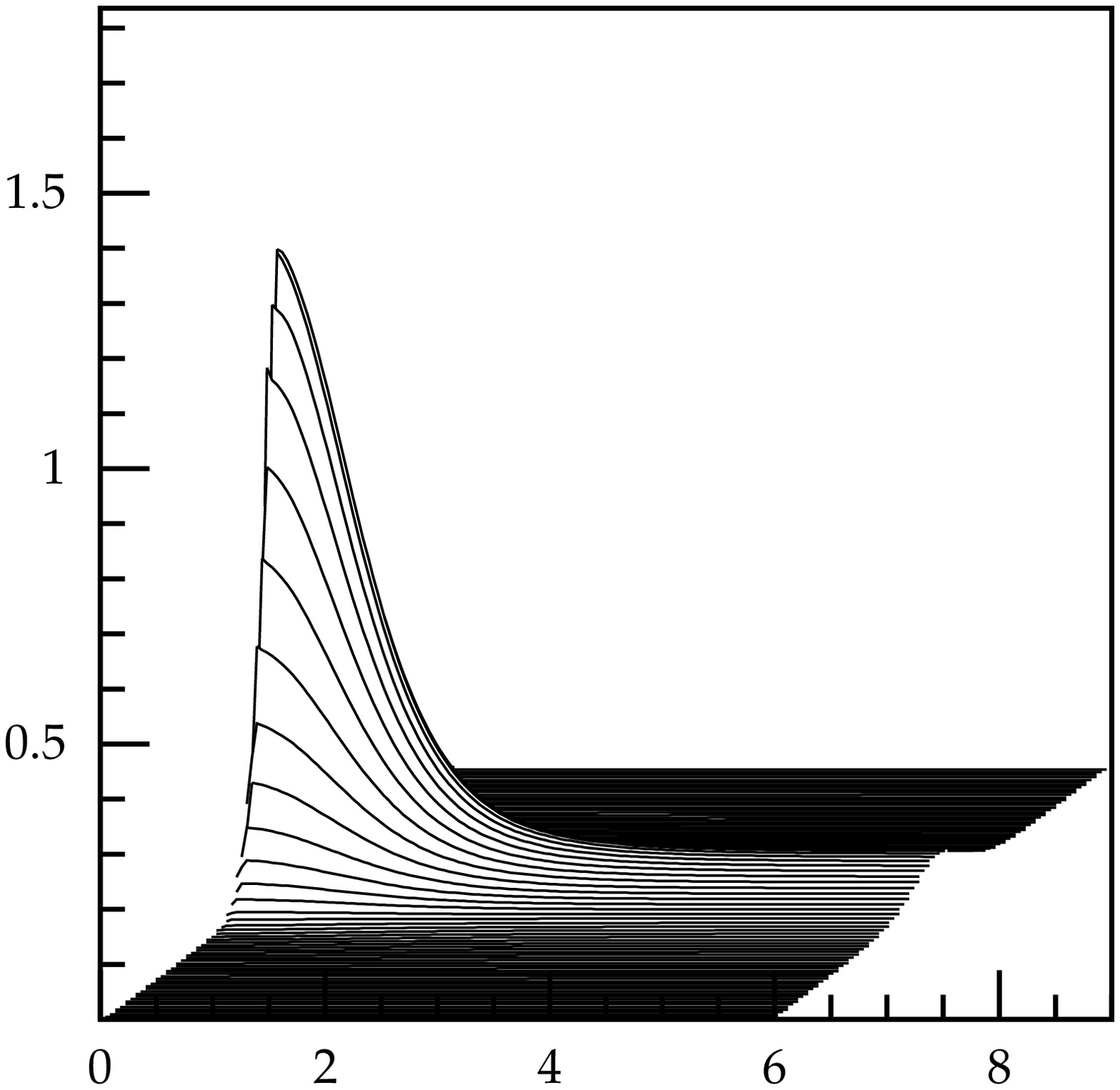}
\end{picture}
\caption{Skyrmion on the Wall ($\delta = 0.3$). 
All expressions are shown using the $r,x$ coordinates:
a) $f$; b) $g$; c) Energy density }
\hfill\newline
\begin{picture}(8,8)
\epsfxsize=8cm
\epsffile{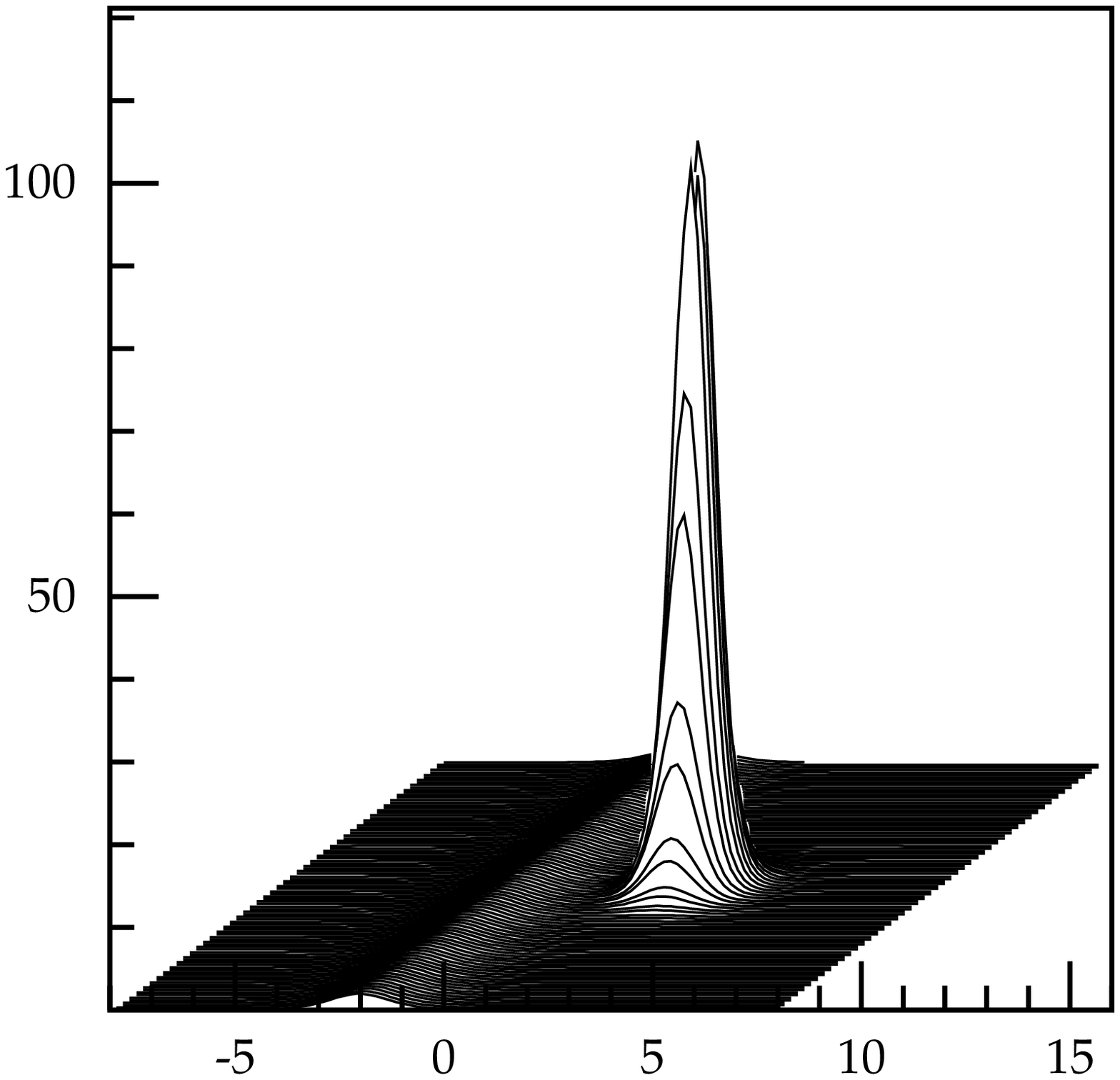}
\end{picture}
\begin{picture}(8,8)
\epsfxsize=8cm
\epsffile{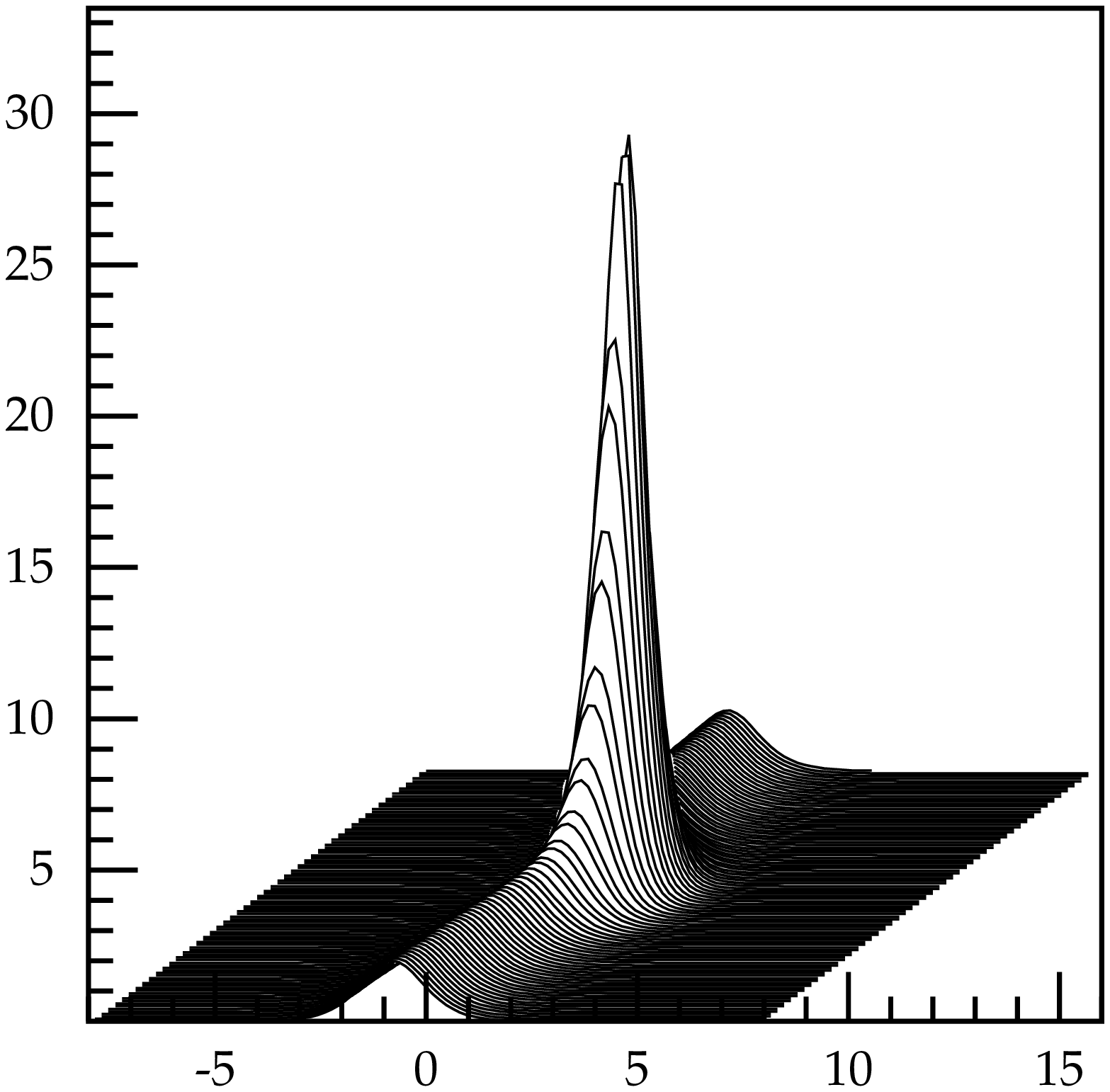}
\end{picture}
\begin{picture}(8,8)
\epsfxsize=8cm
\epsffile{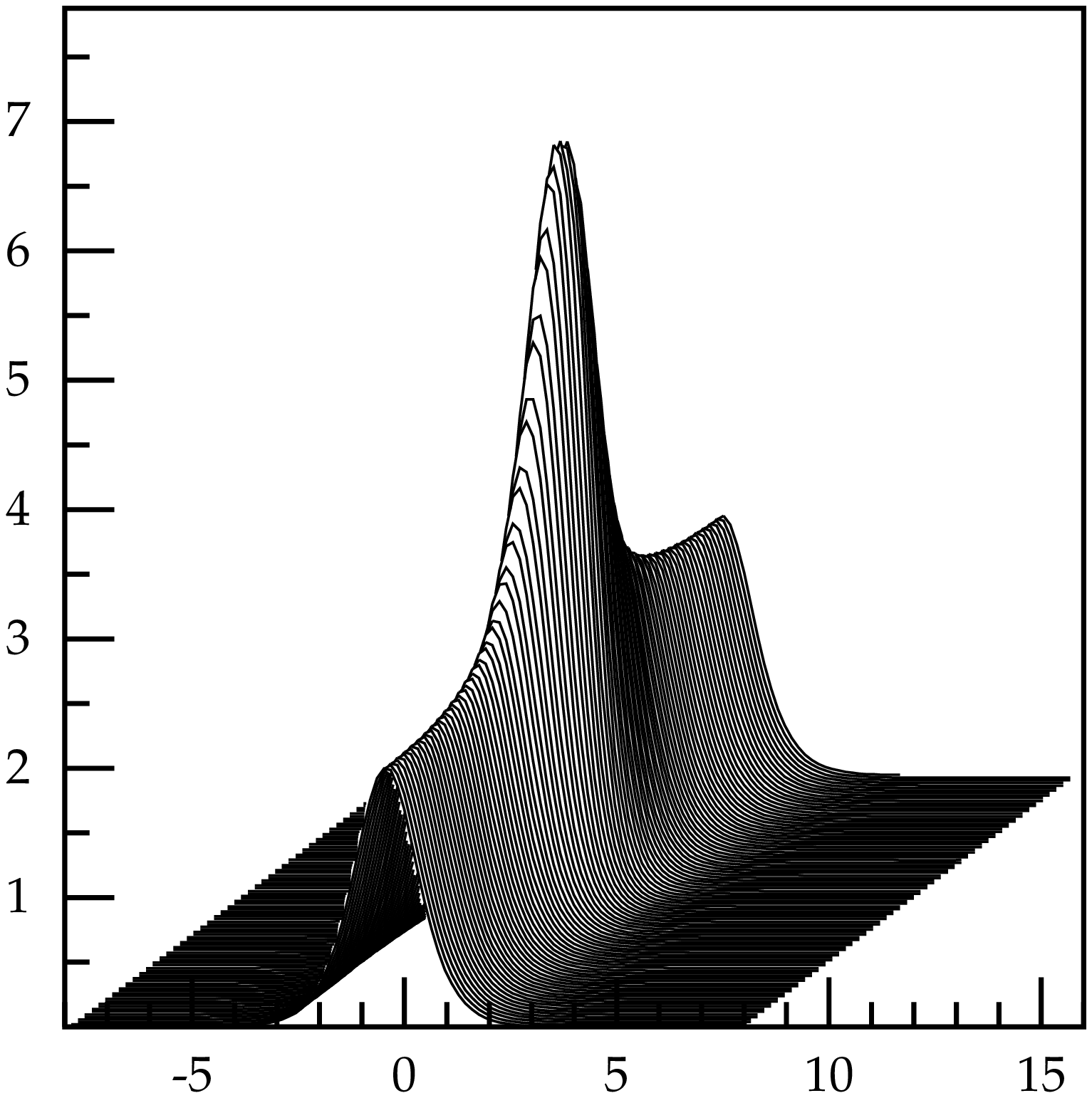}
\end{picture}
\begin{picture}(8,8)
\epsfxsize=8cm
\epsffile{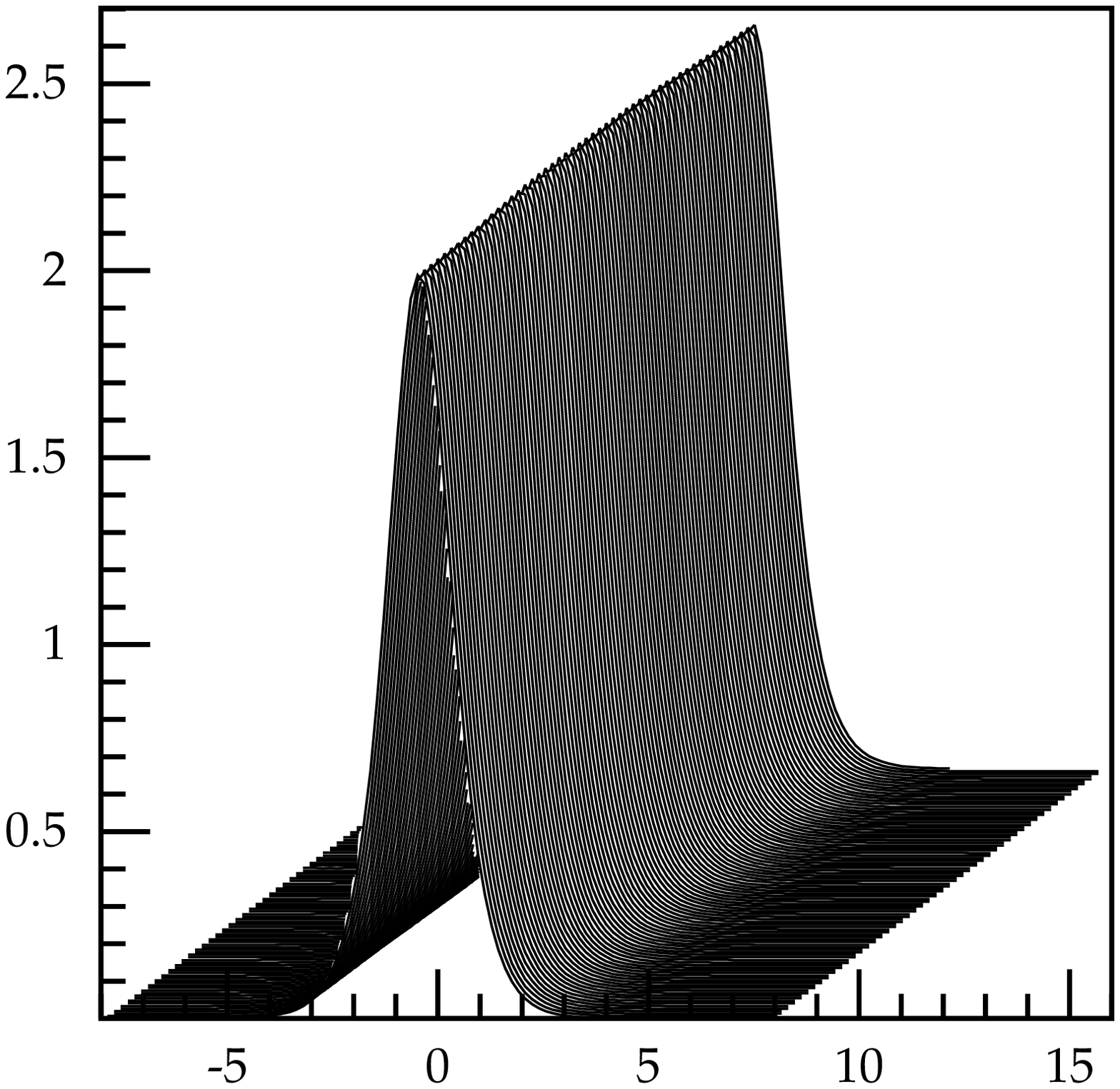}
\end{picture}
\caption{Skyrmion absorption by the Wall ($\delta = 0$). 
Energy density cross section perpendicular to the wall at: 
a) t=0; b) t=37.5; c) t=50; d) t=65 }
\hfill\newline
\begin{picture}(8,8)
\epsfxsize=8cm
\epsffile{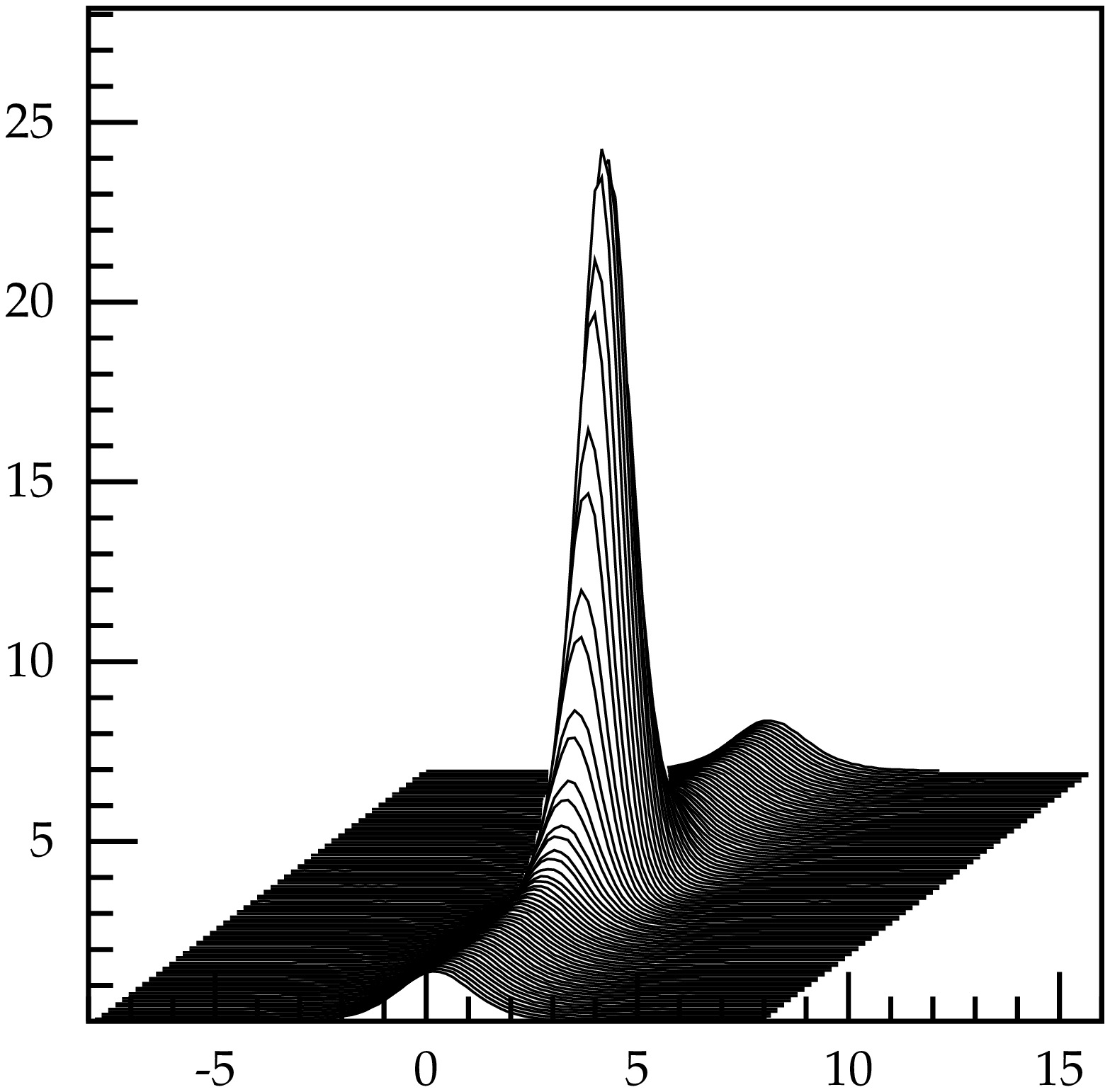}
\end{picture}
\caption{Static Skyrmions on the Wall ($\delta = 0.3$) : cross section of the 
energy density perpendicular to the wall}
\hfill\newline
\begin{picture}(8,8)
\epsfxsize=8cm
\epsffile{SKm2_1sPotEn0.ps}
\end{picture}
\begin{picture}(8,8)
\epsfxsize=8cm
\epsffile{SKm2_1sPotEn75.ps}
\end{picture}
\caption{Two Skyrmions on the Wall ($\delta = 0.3$) : cross section of the 
energy density a) perpendicular to the wall b) parallel to the wall }

\end{figure*}

\end{document}